\documentstyle[12pt]{article}

\def\hybrid{\topmargin -20pt	\oddsidemargin 0pt
	\headheight 0pt	\headsep 0pt
        \textwidth 6.35in
        \textheight 9.65in
	\marginparwidth .875in
	\parskip 5pt plus 1pt	\jot = 1.5ex}

 \catcode`\@=11

\@addtoreset{equation}{subsection}
\@addtoreset{footnote}{section}
\def\theequation{\thesubsection.\arabic{equation}}

\newtoks\@stequation

\def\subequations{\refstepcounter{equation}%
  \edef\@savedequation{\the\c@equation}%
  \@stequation=\expandafter{\theequation}
  \edef\@savedtheequation{\the\@stequation}
  \edef\oldtheequation{\theequation}%
  \setcounter{equation}{0}%
  \def\theequation{\oldtheequation\alph{equation}}}

\def\endsubequations{\setcounter{equation}{\@savedequation}%
  \@stequation=\expandafter{\@savedtheequation}%
  \edef\theequation{\the\@stequation}\global\@ignoretrue
  \vspace*{-12pt} \\}

\def\pr{Phys. Rev. \/}
\def\np{Nucl. Phys. \/}
\def\pl{Phys. Lett. \/}

\hybrid
\def\baselinestretch{1.2}
\catcode`\@=11
\newcount\hour
\newcount\minute
\newtoks\amorpm
\hour=\time\divide\hour by60
\minute=\time{\multiply\hour by60 \global\advance\minute by-\hour}
\edef\standardtime{{\ifnum\hour<12 \global\amorpm={am}%
	\else\global\amorpm={pm}\advance\hour by-12 \fi
	\ifnum\hour=0 \hour=12 \fi
	\number\hour:\ifnum\minute<10 0\fi\number\minute\the\amorpm}}
\edef\militarytime{\number\hour:\ifnum\minute<10 0\fi\number\minute}

\def\draftlabel#1{{\@bsphack\if@filesw {\let\thepage\relax
   \xdef\@gtempa{\write\@auxout{\string
      \newlabel{#1}{{\@currentlabel}{\thepage}}}}}\@gtempa
   \if@nobreak \ifvmode\nobreak\fi\fi\fi\@esphack}
	\gdef\@eqnlabel{#1}}
\def\@eqnlabel{}
\def\@vacuum{}
\def\draftmarginnote#1{\marginpar{\raggedright\scriptsize\tt#1}}

\def\draft{\oddsidemargin -.5truein
	\def\@oddfoot{\sl preliminary draft \hfil
	\rm\thepage\hfil\sl\today\quad\militarytime}
	\let\@evenfoot\@oddfoot	\overfullrule 3pt
	\let\label=\draftlabel
	\let\marginnote=\draftmarginnote
   \def\@eqnnum{(\theequation)\rlap{\kern\marginparsep\tt\@eqnlabel}%
\global\let\@eqnlabel\@vacuum}  }

\catcode`\@=11
\def\section{\@startsection {section}{1}{0pt}{-3.5ex plus -1ex minus
 -.2ex}{2.3ex plus .2ex}{\raggedright\large\bf}}
\catcode`\@=12
\newskip\humongous \humongous=0pt plus 1000pt minus 1000pt

\newif\ifdtup

\def\be{\begin{equation}}
\def\ee{\end{equation}}
\def\ba{\begin{eqnarray}}
\def\ea{\end{eqnarray}}
\def\bs{\begin{subequations}}
\def\es{\end{subequations}}

\def\th{\theta}

\def\tb{\bar{\theta}}

\def\t{\tau}
\def\tb{\bar\tau}
\def\im{{\rm Im}\tau}

\def\l{\lambda}

\def\t{\tau}

\def\R{{\cal{R}}}
\def\Q{{\cal{Q}}}

\def\dslash{{\partial\hspace{-.22cm}/}}
\def\hslash{{\rm H}\hspace{-.28cm}/}
\begin{document}
\renewcommand{\theequation}{\thesection.\arabic{equation}}
\begin{titlepage}
\begin{center}

\hfill CERN-TH/95-172\\
\hfill LPTENS-95/29\\
\hfill hep-th/9507051\\

\vskip .1in

{\large \bf
Infrared-Regulated String Theory and Loop Corrections to Coupling
Constants
\footnote{To appear in the proceedings of the Strings-95 conference
in  Los Angeles, CA, 13-18 March 1995}}
\vskip .2in

{\bf Elias Kiritsis and Costas Kounnas\footnote{On leave from Ecole
Normale Sup\'erieure, 24 rue Lhomond, F-75231, Paris, Cedex 05,
FRANCE.}}\\
\vskip
 .1in

{\em Theory Division, CERN,\\ CH-1211,
Geneva 23, SWITZERLAND} \\

\end{center}

\begin{center} {\bf ABSTRACT } \end{center}
\begin{quotation}\noindent

Exact superstring solutions are constructed in 4-D space-time,
with positive curvature
and non-trivial dilaton and antisymmetric tensor fields. The full
spectrum
of string excitations is derived as a function of moduli fields
$T^{i}$ and the scale $\mu^2=1/(k+2)$ which is induced by the
non-zero background fields.
The spectrum of string excitations has a non-zero mass gap $\mu^2$
and in the weak curvature limit ($\mu$ small), $\mu^2$ plays the role
of a well defined infrared regulator, consistent with modular
invariance, gauge invariance,
supersymmetry and chirality.\\
 The effects of a covariantly constant (chomo)magnetic field $H$, as
well as additional curvature can be derived exactly up to one
string-loop level. Thus,  the one-loop corrections to
all couplings (gravitational, gauge and Yukawas) are
unambiguously computed and are finite both in the UltraViolet and the
InfraRed regime.
These corrections are necessary for quantitative string
superunification predictions at low energies. The one-loop
corrections to the couplings are also found to satisfy
Infrared Flow Equations.\\
Having in our disposal an exact description  which goes beyond the
leading order in the $\alpha'$-expansion or the linearized
approximation in the background fields, we find interesting clues
about
the physics of string theory in strong gravitational and magnetic
fields.
In particular, the nature of gravitational or magnetic instabilities
is studied.

\end{quotation}
\vskip 0.5cm
CERN-TH/95-172 \\
July 1995\\
\end{titlepage}
\vfill
\eject
\def\baselinestretch{1.2}
\baselineskip 16 pt
\noindent
\section{Introduction}
\setcounter{equation}{0}

The low energy  properties of the four dimensional $N=1$ superstrings
\cite{cand}-\cite{gepner} are described  by  a special class of $N=1$
supergravity theories \cite{effcl}-\cite{ant}, in which all
interactions are unified.  A sub-class of them seems to extend
successfully
the validity of the supersymmetric standard model up to the string
scale,
$M_{str} \sim {\cal O}( 10^{17} )~$GeV.
At this energy scale,
however,
the first excited string states become important and thus the whole
effective low energy field theory picture breaks
down \cite{n4kounnas}-\cite{kktopol}.
The string unification does not include  only the gauge interactions,
as it happens in conventional grand unified theories, but also the
Yukawa interactions as well as the interactions among the scalars.
 At energies of  order of $M_{str}$,  string unification relations
look similar to those of a conventional supersymmetric grand-unified
field  theory (susy-GUT),

\begin{equation}
\frac{k_i}{\alpha_i(M_{str})}={1\over \alpha_{str}}
\label{shu 1}
\end{equation}

In susy-GUTs, the normalization constants $k_i$ are fixed $only$ for
the
gauge couplings ($k_1=k_2=k_3=1$, $k_{em}=\frac{3}{8}$), but there
are no relations among gauge and Yukawa couplings at all. In string
effective theories, however, the normalization constants ($k_i$) are
known for both gauge and Yukawa interactions. The above unification
relations
among the couplings are corrected at the quantum level not only by
the conventional field theory renormalizable interactions involving
light-mass states but also by string corrections involving both the
light and the infinite tower of string massive  states. It is of main
importance that the superstring corrections are finite in the
ultraviolet regime and thus one expects to obtain all quantum
corrections without ambiguities. In particular one expects
the string unification predictions at a scale $\mu<M_{str}$ to have
the following form,

\begin{equation}
\frac{k_i}{\alpha_i(\mu)}={1\over \alpha_{str}}+
\frac{b_{i}}{4\pi}\log \frac{\mu^{2}}{M^{2}_{str}}+\Delta_{i}(T^{A}).
\label{shu}
\end{equation}

The logarithmic behavior in the above formula is due to the light
states with masses lower than the scale $\mu$ and $\Delta_{i}(T^{A})$
are calculable, $finite$ quantities for any particular string
solution. Thus, the predictability of a given string solution is
extended to all low energy coupling constants once the string-induced
corrections are determined. $\beta$-functions in string theory were
calculated first in \cite{min}

It turns out that $\Delta_{i}(T^A)$ are non-trivial functions of the
vacuum
expectation values  of  some gauge singlet fields
\cite{moduli,dfkz,ant,n2}, the so-called moduli\footnote{The moduli
fields
are flat directions at the string classical level and they remain
flat in
string perturbation theory, in the exact supersymmetry limit.}.
The $\Delta_{i}(T^A)$ are target space duality-invariant functions,
which depend on the particular
string ground state. Several results for $\Delta_{i}(T^A)$ exist by
now \cite{moduli,dfkz,ant,n2} in the exact supersymmetric limit, in
many string
solutions based on orbifold \cite{orbifold} and fermionic
constructions \cite{abk4d}. As we will see later, $\Delta_{i}(T^A)$
are in principle well defined calculable quantities once we
perform our calculations  at the string level where all interactions
including gravity are  consistently defined.

Although at this stage we do not know the details of supersymmetry
breaking,
we should stress here that the corrections to dimensionless coupling
constants
(e.g. the coefficients of dimension four operators) are still exact
if the low energy scale  $\mu$ is chosen above the threshold of
supersymmetric partners of light states, ($m_{susy}\le \mu$). This is
due to the fact that the soft breaking terms in the effective theory
do not affect the  renormalizations properties of the dimensionless
couplings.
For the corrections and the structure of the soft breaking parameters
only qualitative results exist up to now although this is a subject
of an intensive study.

In the past, there was an obstruction in determining  the exact form
of the string corrections $\Delta_{i}(T^A)$ due to the infrared
divergences of the on-shell calculations in string theory.
In a second quantized field theory, we can avoid the IR-divergences
due to the massless  particles using off-shell calculations.
In string theory we cannot do this since string field theory is not
very useful
computationally up to now and  in the first quantized formulation,
which is available at present, we do not know how to go consistently
off-shell.
Even in field theory there are problems in defining an infrared
regulator for chiral fermions especially in the presence of
space-time supersymmetry.

The idea we will employ here is to modify slightly the ground state
of interest in string theory
so that it develops a mass gap. It is known already in field theory
that a space of negative curvature provides fields (scalars, fermions
 vectors etc.) with such a mass gap.
We will see however that string theory contains the fields (namely
the antisymmetric tensor) which when they acquire some suitable
expectation values they can provide a mass gap for chiral fermions
without running into trouble with anomalies.

Let us indicate here how an expectation value for the dilaton can
give masses to bosonic fields.
The dilaton couples generically to (massless) bosonic fields in a
universal fashion:

\be
S_{T}=\int e^{-2\Phi}\partial_{\mu}T\partial^{\mu}T
\ee
where we considered the case of a scalar field $T$.
To find the spectrum of the fluctuations of T we have to define
$\tilde T=e^{-\Phi}T$ so that kinetic terms are diagonalized. Then,
the action becomes
\be
S_{T}=\int \partial_{\mu}\tilde T\partial^{\mu}\tilde
T+\left[\partial_{\mu}\Phi\partial^{\mu}\Phi-\partial_{\mu}
\partial^{\mu}\Phi\right]\tilde T^2
\ee
It is obvious that if $\langle \Phi\rangle =Q_{\mu}x^{\mu}$ then the
scalar
$\tilde T$ acquires a mass$^2 ~Q_{\mu}Q^{\mu}$\footnote{This was
observed in \cite{aben} with
$Q_{\mu}$ being timelike.} which is positive when
$Q_{\mu}$ is spacelike.
Similar remarks apply to higher spin bosonic fields.
This mechanism via the dilaton cannot give masses to fermions since
the extra shift obtained by the diagonalization is a total
divergence.

Consider a chiral fermion with its universal coupling to the
antisymmetric tensor:

\be
S_{\psi}=\int \bar \psi [{\buildrel{\leftrightarrow}\over
\dslash}+{\hslash}]\psi
\ee
where $H_{\mu}={\epsilon_{\mu}}^{\nu\rho\sigma}H_{\nu\rho\sigma}$ is
the dual
of the field strength of the antisymmetric tensor.
If $\langle H_{\mu}\rangle =Q_{\mu}$, then the Dirac operator
acquires
a mass gap proportional to $Q_{\mu}Q^{\mu}$.

Thus we need to find exact string ground states (CFTs) which
implement the mechanism sketched above.

In particular we would like our background to have the following
properties:

{\bf 1.} The string spectrum must have a mass gap $\mu^2$.
          In particular, chiral fermions should be regulated
consistently.

{\bf 2.} We should be able to take the limit $\mu^2\to 0$.

{\bf 3.} It should have $c=(6,4)$ (in the heterotic case) so that it
can be coupled to any
         internal CFT with $c=(9,22)$.

{\bf 4.} It should preserve as many spacetime supersymmetries of the
original theory, as possible.

{\bf 5.} We should be able to calculate the regulated quantities
relevant for
the effective field theory.

{\bf 6.} Vertices for spacetime fields (like $F_{\mu\nu}^{a}$) should
be
          well defined operators on the world-sheet.

{\bf 7.} The theory should be modular invariant (which guarantees the
absence
of anomalies).

{\bf 8.} Such a regularization should be possible also at the
effective field theory level. In this way, calculations in the
fundamental theory can be matched
without any ambiguity to those of the effective field theory.

CFTs with the properties above employ special four-dimensional spaces
as
superconformal building blocks with
${\hat c}=4$ and $N=4$ superconformal
symmetry \cite{n4kounnas,worm}. The full spectrum of string
excitations for the superstring solutions based on those
four-dimensional subspaces, can be derived using the techniques
developed in \cite{worm}. The spectrum does have a mass gap, which is
proportional
to the curvature of the non-trivial  four-dimensional spacetime.
Comparing the spectrum in  a flat background with that in curved
space, we observe a shifting of all massless states by an amount
proportional to the spacetime curvature, $\Delta m^2=Q^2=\mu^2$,
where $Q$ is the Liouville background charge and $\mu$ is the IR
cutoff. In particular, all gauge symmetries of the original vacuum
are spontaneously broken\footnote{This is not the usual Higgs type of
breaking.
Gauge symmetry is spontaneously broken here by non-trivial
expectation values of fields in the gravitational sector.}.
What is
also interesting is that the shifted spectrum in curved space is
equal for bosons and fermions due to the existence of a new
space-time supersymmetry defined in curved spacetime
\cite{n4kounnas,worm}. Therefore, our curved spacetime
infrared
regularization is consistent with supersymmetry, and can be
used either in field theory or string theory.

Once we regulate the IR, the one-loop corrections to the couplings
can be calculated using the background field method.
We will turn on, (chromo)magnetic fields as well as curvature and
calculate their
effective action at one-loop from which the coupling corrections can
be read directly. Of course, since we work in the first quantized
formulation the background gauge and gravitational fields have to
satisfy the string equations of motion. It turns out that in the IR
regulated string theory there are marginal perturbations which turn
on precisely the background fields we need.
This provides a new class of exact magnetic field solutions to closed
strings\footnote{Electromagnetic backgrounds in open strings have
been discussed in \cite{open}. Magnetic  backgrounds in closed string
theory have
been discussed in \cite{bk,rt}.}.
As a byproduct we obtain the exact spectrum of heterotic strings in
the presence of such (chromo)magnetic fields.

Finite magnetic fields generically break the spacetime
supersymmetries\footnote{ Internal magnetic fields with emphasis on
supersymmetry breaking were discussed recently in \cite{bachas}.
Also the stringy Scherk-Schwarz type of breaking, \cite{SS} has a
similar interpretation.}. We
will analyze the presence of tachyonic instabilities induced by such
magnetic fields.
First, we find \cite{magn} that unlike the field theory case, we have
a maximum value for the magnetic field
\be
H_{\rm max}={M^2_{\rm string}\over \sqrt{2}}
\ee
where, as usual, $M^2_{\rm string}=1/\alpha'$.
When $H\to H_{\rm max}$, all states that couple to the
magnetic fields (that is, they are either charged or
have angular momentum) become infinitely massive and
decouple from the theory.
This signals the presence of a boundary in the moduli space of the
magnetic field.

In field theory there is a critical magnetic field
\be
H^{\rm field\;\;theory}_{\rm crit}\sim {\cal O}(\mu^2)
\ee
beyond which the theory is unstable. Here $\mu$ is the mass gap of
the theory in the charged sector.
In the string case there is also a lower critical  magnetic field
beyond which the theory is unstable but it scales differently
\be
H^{\rm string\;\;theory}_{\rm lower\;\;crit}\sim {\cal O}(\mu M_{\rm
string})
\ee
where again $\mu$ is the mass gap.
This difference is due to the different breaking of gauge symmetry.
In our string solutions the gauge symmetries are broken by
expectation values of
graviton, antisymmetric tensor and dilaton rather than Higgs fields.

In string theory the spectrum is a non-linear function of the
magnetic field due to the gravitational backreaction.
The effect of the non-linearity is that there is also an upper
critical magnetic field $H^{\rm crit}_{\rm upper}$ such that $H_{\rm
max}-H^{\rm crit}_{\rm upper}\sim {\cal O}(\mu M_{str})$ so that the
theory is stable for
\be
H^{\rm crit}_{\rm upper}\leq H\leq H_{\rm max}
\ee

Similar remarks apply to curvature perturbations. Again, there are
tachyonic instabilities due to the breaking of spacetime
supersymmetry for a region of the parameters.

Most of the work presented here has already appeared in
\cite{magn,trieste}
We present also some new results in section 5.

\section{The IR regulated String Theory}
\setcounter{equation}{0}

We will choose the 4-D CFT which will replace flat space to
correspond to the W-space described by
the $SU(2)_{k}\otimes R_{Q}$ model.
It contains a non-compact direction with a linear dilaton
$\Phi=Qx^{0}$
as well as the $SU(2)_{k}$ WZW model.
Q is related to $k$ as $Q=1/\sqrt{k+2}$ so that the CFT has the same
central charge as flat space.
We will define $\mu^2=1/(k+2)$, $\mu$ is directly related to the mass
gap of the regulated theory.
The GSO projection couples the SU(2) spin with the spacetime helicity
\cite{trieste}.
This effectively projects out the half-integral spins and replaces
$SU(2)$ with $SO(3)$. $k$ should be an even positive integer for
consistency.
For any ground state of the heterotic string with $N<4$ spacetime
supersymmetry
the regulated vacuum amplitude turns out to be

\be
Z(\mu)={1\over V(\mu)}\Gamma_{0}(\mu)Z_{0}\label{21}
\ee
where $V(\mu)=1/8\pi\mu^3$ is the volume of the nontrivial background
and $Z_{0}$ is the vacuum amplitude for the unregulated theory, which
can be written as
\be
Z_{0}(\t,\tb)={1\over \im|\eta|^4}\sum_{a,b=0}^{1}{\th[^a_b]\over
\eta}
C[^a_b](\t,\tb)
\label{22}
\ee
where we have separated the generic 4-d contribution. The factor
$C[^a_b]$ is the trace in the $(^a_b)$ sector of the internal CFT.
Finally, $\Gamma_{0}(\mu)$ is proportional to the $SO(3)_{k/2}$
partition function
\be
\Gamma_{0}(\mu)={1\over 2}[({\rm Im}\tau)^{\frac{1}{2}}
\eta(\tau){\bar
\eta}({\bar \tau})]^{3}~~\sum_{a,b=0}^{1}e^{-i\pi
kab/2}\sum_{l=0}^{k}e^{i\pi
bl}\chi_{l}(\t)\bar \chi_{(1-2a)l+ak}(\tb)
\label{23}
\ee
where $\chi_{l}$ are the standard $SU(2)_{k}$ characters.
We have also the correct limit $Z(\mu)\to Z_{0}$ as $\mu\to 0$.

There is a simple expression for $\Gamma_{0}(\mu)$
\be
\Gamma_{0}(\mu)=-{1\over 2\pi}X'(\mu)
\label{24}
\ee
where prime stands for derivative with respect to $\mu^2$
and
\be
X(\mu)={1\over 2\mu}\sum_{m,n\in
Z}e^{i\pi(m+n+mn)}\;\exp\left[-{\pi\over
4\mu^2\im}|m-n\t|^2\right]=\sqrt{\im}\sum_{m,n\in Z}e^{i\pi
n}q^{{1\over
4}Q^{2}_{L}}\bar q^{{1\over 4} Q^{2}_{R}}
\label{25}
\ee
with
\be
Q_{L}=2\mu\left(m-{n+1\over 2}\right)+{n\over
2\mu}\;\;,\;\;Q_{R}=2\mu\left(m-{n+1\over 2}\right)-{n\over 2\mu}
\label{26}
\ee
It can be also written in terms of the usual torroidal sum
\cite{trieste}:
\be
X(\mu)=Z_{T}(\mu)-Z_{T}(2\mu)
\ee
\be
Z_{T}(\mu)=Z_{T}(1/\mu)=\sqrt{\im}\sum_{m,n\in Z}q^{{1\over
4}\left(m\mu+n/\mu\right)^2}
\bar q^{{1\over 4}\left(m\mu-n/\mu\right)^2}\label{37}
\ee
Note that $X(\mu)$ is modular invariant.

The leading infrared behavior can be read from (\ref{24}),
(\ref{25})
to be
\be
Z(\mu)\to \sqrt{\im}e^{-\pi\im\mu^2}
\label{27}
\ee
as $\im\to \infty$ that indicates explicitly the presence of the mass
gap.

More details on this theory can be found in \cite{magn,trieste}.

\section{Non-zero ${\bf F^{a}_{\mu\nu}}$ and  $
R_{\mu\nu}^{\rho\sigma}$ Background in Superstrings}
\setcounter{equation}{0}

As mentioned in the introduction, in order to calculate the
renormalization of the effective couplings we need to turn on
backgrounds for gauge and gravitational fields.
Thus, our aim is to define the  deformation of the two-dimensional
superconformal theory  which corresponds to a non-zero field strength
$F^{a}_{\mu\nu}$ and $R_{\mu\nu\rho\sigma}$
background
and find the integrated  one-loop
partition function $ Z(\mu,F,\R)$,  where $F$ is related to  the
magnitude
of the field strength,
$F^2 \sim \langle F^{a}_{\mu\nu}F_{a}^{\mu\nu}\rangle$ and $\R$ is
that of the curvature,  $\langle
R_{\mu\nu\rho\sigma}R^{\mu\nu\rho\sigma}\rangle\sim \R^2$.

\begin{equation}
Z[\mu,F_i,\R]=\frac{1}{V(\mu)} \int_{\cal F}
\frac{ d\tau d{\bar\tau} }{ ({\rm Im}\tau)^2 }
Z[\mu,F_i,\R;\tau,{\bar\tau}]
\label{intpart}
\end{equation}
The index $i$ labels different simple or $U(1)$ factors of the gauge
group of the ground state.

 In flat space, a small non-zero  $F_{\mu\nu}^a$ background gives
rise
to an infinitesimal deformation  of the 2-d $\sigma$-model action
given by,
\begin{equation}
\Delta S^{2d}_{\rm flat}=\int dzd{\bar z}\;F_{\mu\nu}^a[x^{\mu}
\partial_z x^{\nu}+\psi^{\mu}\psi^{\nu}]{\bar J}^a
\label{fdef}
\end{equation}
Observe that for $F^a_{\mu\nu}$ constant (constant magnetic field),
the left moving operator $[x^{\mu} \partial_z
x^{\nu}+\psi^{\mu}\psi^{\nu}]$ is not a well-defined $(1,0)$ operator
on the world sheet. Even though  the right moving Kac-Moody current
${\bar J}_a$ is a well-defined $(0,1)$ operator, the total
deformation
is not integrable in flat space. Indeed, the 2-d $\sigma$-model
$\beta$-functions are not satisfied in the presence of a constant
magnetic field. This follows from the fact that there is a
$non$-$trivial$ $back$-$reaction$ on the gravitational background due
the non-zero
magnetic field.
\def\q{{\cal Q}}
\def\pq{{\bar {\cal P}}}
\def\rt{{\cal R}}

In the W-space, however, the vertex operator which turns on a
(chromo)magnetic field background $B^{a}_{i}$ is
\be
V^{a}_{i}=(J^i+{1\over 2}\epsilon^{i,j,k}\psi^j\psi^k)\bar J^{a}
\ee
This vertex operator is of the current-current type. In order for
such perturbations to be marginal (equivalently the background to
satisfy the string equations of motion) we need to pick a single
index $i$, which we choose to be $i=3$ and need to restrict the gauge
group index $a$ to be in the Cartan of the
gauge group.
We will normalize the antiholomorphic currents $\bar J^a_i$ in each
simple or U(1) component $G_{i}$ of the gauge group $G$ as
\be
\langle \bar J^a_i(\bar z)\bar J^b_j(0)\rangle
={k_{i}\delta^{ij}\over 2}{\delta^{ab}\over \bar z^2}\label{norm}
\ee
With this normalization, the field theory gauge coupling is
$g_{i}^2=2/k_{i}$.
Thus the most general marginal (chromo)magnetic field is generated
from the following vertex operator
\be
V_{magn}={(J^3+\psi^1\psi^2)\over \sqrt{k+2}}{{\vec F_{i}}\cdot{\bar
{\vec J_{i}}}\over \sqrt{k_{i}}}\label{magn}
\ee
where the index $i$ labels the simple or $U(1)$ components $G_{i}$ of
the gauge group and $\bar{\vec J_{i}}$ is a $r_{i}$-dimensional
vector of currents in the Cartan of the group
$G_{i}$ ($r_{i}$ is the rank of $G_{i}$).
The repeated index $i$ implies summation over the simple components
of the gauge group.

We would like to obtain the exact one-loop partition function in the
presence of such perturbation. Since this is an abelian
current-current perturbation,
the deformed partition function can be obtained by an $O(1,N)$, boost
($N=\sum_{i}r_{i}$)
of the charged lattice of the undeformed partition function, computed
in the previous section.

We will indicate the method in the case where we turn on a single
magnetic field $F$, in a gauge group factor with central element
$k_{g}$, in which case
\be
V_{F}= F{(J^3+\psi^1\psi^2)\over \sqrt{k+2}}{\bar J\over
\sqrt{k_{g}}}
\ee
Let us denote by $\q$ the zero mode of the holomorphic helicity
current
$\psi^1\psi^2$,  $\pq$ the zero mode of the antiholomorphic current
$\bar J$ and $I,\bar I$ the zero modes of the $SU(2)$ currents
$J^{3},\bar J^{3}$ respectively.
Then, the relevant parts of $L_{0}$ and $\bar L_{0}$ are
\be
L_{0}={\q^2\over 2}+{I^2\over k}+\cdots\;\;\;,\;\;\;\bar
L_{0}={\pq^2\over k_{g}}+\cdots\label{k1}
\ee
We will rewrite $L_{0}$ as
\be
L_{0}={(\q+I)^2\over k+2}+{k\over 2(k+2)}\left(\q-{2\over
k}I\right)^2+\cdots
\ee
where we have separated the relevant supersymmetric zero mode $\q+I$
and its
orthogonal complement $\q-2I/k$ which will be a neutral spectator to
the perturbing process.
What remains to be done is an $O(1,1)$ boost that mixes the
holomorphic current
$\q+I$ and the antiholomorphic one $\pq$.
This is straighforward with the result
\be
L_{0}'= {k\over 2(k+2)}\left(\q-{2\over k}I\right)^2+\left(\cosh x
{\q+I\over \sqrt{k+2}}+\sinh x {\pq\over
\sqrt{k_{g}}}\right)^2+\cdots
\label{k2}
\ee
\be
\bar L_{0}'= \left(\sinh x {\q+I\over \sqrt{k+2}}+\cosh x {\pq\over
\sqrt{k_{g}}}\right)^2+\cdots\label{k3}
\ee
where $x$ is the parameter of the $O(1,1)$ boost.
Thus we obtain from (\ref{k2}), (\ref{k3}) the change of $L_{0}$,
$\bar L_{0}$ as
\be
\delta L_{0}\equiv L_{0}'-L_{0}=\delta \bar L_{0}\equiv   \bar
L_{0}'-\bar L_{0}=
F{(\q+I)\over \sqrt{k+2}}{\pq\over \sqrt{k_{g}}}+{\sqrt{1+F^2}-1\over
2}\left[
{(\q+I)^2\over k+2}+{\pq^2\over k_{g}}\right]
\ee
where we have identified
\be
F\equiv \sinh (2x)
\ee

We are now able to compute with the more general marginal
perturbation
which is a sum of the general magnetic perturbation (\ref{magn})
and the gravitational perturbation
\be
V_{grav}= \rt {(J^3+\psi^1\psi^2)\over \sqrt{k+2}}{\bar J^3\over
\sqrt{k}}
\ee
The only extra ingredient we need is an $O(1+N)$ transformation to
mix
the antiholomorphic currents.
Thus, we obtain
\be
\delta L_{0}=\delta \bar L_{0}=\left[{\rt\bar I\over \sqrt{k}}+{{\vec
F}_{i}\cdot {\bar {\vec {\cal P}}}_{i}\over
\sqrt{k_{i}}}\right]{(\q+I)\over \sqrt{k+2}}+\label{gener}
\ee
$$+{\sqrt{1+\rt^2+{\vec F}_{i}\cdot {\vec F}_{i}}-1\over 2}\left[
{(\q+I)^2\over k+2}+(\rt^2+{\vec F}_{i}\cdot {\vec F}_{i})^{-1}\left(
{\rt\bar I\over \sqrt{k}}+{{\vec F}_{i}\cdot {\bar {\vec {\cal
P}}}_{i}\over \sqrt{k_{i}}}\right)^2\right]
$$

{}From now on we focus in the case where we have a single
(chromo)magnetic field $F$ as well as the curvature perturbation
$\rt$.
Then (\ref{gener}) simplifies to

\be
\delta L_{0}=\delta \bar L_{0}=
\left[\rt{\bar I\over \sqrt{k}}+F{\pq\over
\sqrt{k_{g}}}\right]{(\q+I)\over \sqrt{k+2}}+\label{gener2}
\ee
$$+{\sqrt{1+\rt^2+F^2}-1\over 2}\left[
{(\q+I)^2\over k+2}+(\rt^2+F^2)^{-1}\left(
\rt{\bar I\over \sqrt{k}}+F{\pq\over \sqrt{k_{i}}}\right)^2\right]
$$

Eq. (\ref{gener2}) can be written in the following form which will be
useful
in order to compare with the field theory limit
\be
\delta L_{0}={1+\sqrt{1+F^2+{\cal R}^2}\over 2}\left[{(\q+I)\over
\sqrt{k+2}}+{1\over 1+\sqrt{1+F^2+{\cal R}^2}}\left(\rt{\bar I\over
\sqrt{k}}
+F{\pq\over \sqrt{k_{g}}}\right)\right]^2
\ee
$$-{(\q+I)^2\over k+2}
$$
and for $\rt=0$ as
\be
\delta L_{0}={1+\sqrt{1+F^2}\over 2}\left[{(\q+I)\over
\sqrt{k+2}}+{F\over 1+\sqrt{1+F^2}}{\pq\over \sqrt{k_{g}}}\right]^2
-{(\q+I)^2\over k+2}\label{k10}
\ee

Eq. (\ref{21}) along with (\ref{gener}) provide the complete and
exact spectrum of string theory in the presence of the
(chromo)magnetic
fields $\vec F_{i}$ and curvature $\rt$.
We will analyse first the case of a single magnetic field $F$ and use
(\ref{k10}). Since for physical states $L_{0}=\bar L_{0}$ it is
enough to look
at $L_{0}=M_{L}^2$ which in our conventions is the side that has
$N=1$ superconformal symmetry.

\be
M^2_{L}=-{1\over 2}+{\q^2\over 2}+{1\over
2}\sum_{i=1}^{3}\q_{i}^2+{(j+1/2)^2-(\q+I)^2\over
k+2}+E_{0}+\label{k11}
\ee
$${1+\sqrt{1+F^2}\over 2}\left[{(\q+I)\over \sqrt{k+2}}+{F\over
1+\sqrt{1+F^2}}{\pq\over \sqrt{k_{g}}}\right]^2
$$
where, the $-1/2$ is the universal intercept in the N=1 side,
$\q_{i}$ are the internal helicity operators (associated to the
internal left-moving fermions),
$E_{0}$ contains the oscillator contributions as well as the internal
lattice
(or twisted) contributions, and
$j=0,1,2,\cdots,k/2$\footnote{Remember that $k$ is an even integer
for $SO(3)$.}, $j\geq |I|\in Z$.
We can see already a reason here for the need of the SO(3)
projection. We do not want half-integral values of $I$ to change the
half-integrality of the spacetime helicity $\q$.

Let us look first at how the low lying spectrum of space-time
fermions
is modified.
For this we have to take $\q=\q_{i}=\pm 1/2$.
Then $M_{L}^2$ can be written as a sum of positive factors,
$E_{0}\geq 0$,
$(j+1/2)^2\geq (\pm 1/2+I)^2$ and
\be
 {1+\sqrt{1+F^2}\over 2}\left[{(\q+I)\over \sqrt{k+2}}+{F\over
1+\sqrt{1+F^2}}{\pq\over \sqrt{k_{g}}}\right]^2
\geq 0\label{k14}
\ee
Thus fermions cannot become tachyonic and this a good consistency
check for our spectrum since a ``tachyonic" fermion is a ghost.
This argument can be generalized to all spacetime fermions in the
theory.

Bosonic states can become tachyonic though, but for this to happen,
as in field theory they need to have non-zero helicity.
It can be shown that for $k$ positive only $|\q|=1$,
$j=|I|=0,1,2,\cdots,k/2$ states can become tachyonic\footnote{This is
unlike the case of \cite{rt} where states with higher helicities
become tachyonic.}.

By also imposing $L_{0}=\bar L_{0}$ we obtain
\be
\q^2-{2\over k_{g}}\pq^2+1\geq 0\label{k16}
\ee
and thus the minimal value for $M_{L}^2$ can be written as
\be
M^2_{min}={\q^2-1\over 2}+{(|I|+1/2)^2-(\q+I)^2\over k+2}+
{1+\sqrt{1+F^2}\over 2}\left[{(\q+I)\over \sqrt{k+2}}+{F\over
1+\sqrt{1+F^2}}{\pq\over \sqrt{k_{g}}}\right]^2
\label{k13}
\ee
Let us introduce the variables
\be
H={F\over \sqrt{2}(1+\sqrt{1+F^2})}\;\;\;,\;\;\;e=\sqrt{2\over
k_{g}}\pq
\ee
$H$ is the natural magnetic field from the $\sigma$-model point of
view \cite{magn} and $e$ is the charge.
Notice that while $F$ varies along the whole real line, $|H|\leq
1/\sqrt{2}$.
{}From (\ref{k16})
\be
e^2\leq \q^2+1\label{k17}
\ee
Then, there are tachyons provided
\be
{1\over 1-2H^2}\left({(\q+I)\over
\sqrt{k+2}}+eH\right)^2+{\q^2-1\over 2}+{(|I|+1/2)^2-(\q+I)^2\over
k+2}\leq 0\label{tach}
\ee
In fact it is not difficult to see that the first instability appears
due to
$I=0$ states becoming tachyonic.
We will leave the charge free for the moment, although there are
certainly constraints on it, depending on the gauge group.
For example for the $E_{6}$ or $E_{8}$ groups we have
$e^2_{min}=1/4$, and for all realistic non-abelian gauge groups
$e_{min}={\cal O}(1)$.
For torroidal $U(1)$'s however $e_{min}$ can become arbitrarily small
by tuning the parameters of the torus. Note however that in any case
for the potential tachyonic states with $|\q|=1$ the charge must
satisfy
\be
{1\over 2(k+2)} \leq e^2 \leq 2\label{k20}
\ee

Thus for $|\q|=1$ we obtain the presence of tachyons provided that
\be
H^{\rm crit}_{\rm min}\leq |H| \leq H^{\rm crit}_{\rm max}
\ee
with
\be
H^{\rm crit}_{\rm min}={\mu\over |e|} {1-{\sqrt{3}\over
2}\sqrt{1-{1\over 2}\left({\mu\over e}\right)^2}\over
1+{3\over 2}\left({\mu\over e}\right)^2}
\ee
\be
H^{\rm crit}_{\rm max}={\mu\over |e|}{J+1+ \sqrt{\left(J+{3\over
4}\right)\left(1-2\left(J+{1\over 2}\right)^2{\mu^2\over
e^2}\right)}\over
1+\left(2J+{3\over 2}\right){\mu^2\over e^2}}
\label{q1}
\ee
where
\be
J={\rm integral ~~part~~of~~}-{1\over 2}+{|e|\over \sqrt{2}\mu}
\ee
We have also introduced the IR cutoff scale $k+2=1/\mu^2$.

We note that for small $\mu$ and $|e|\sim {\cal O}(1)$ $H^{\rm
crit}_{\rm min}$ is of order ${\cal O}(\mu)$.
However $H^{\rm crit}_{\rm max}$ is below $H_{\rm
max}=1/\sqrt{2}$\footnote{We will frequently use dimensionless
notation, $\alpha'=1$. Dimensions can be easily reinstated.} by
an amount of order ${\cal O}(\mu)$.
Thus for small values of $H$ there are no tachyons until a critical
value $H^{\rm crit}_{\rm min}$ where the theory becomes unstable. For
$|H|\geq H^{\rm crit}_{\rm max}$ the theory is stable again till the
boundary $H=1/\sqrt{2}$.
It is interesting to note that if there is a charge in the theory
with the value $|e|=\sqrt{2}\mu$ then $H^{\rm crit}_{\rm
max}=1/\sqrt{2}$ so there
is no region
of stability for large magnetic fields.
For small $\mu$ there are always charges satisfying (\ref{k20}) which
implies that there is always a magnetic instability.
However even for $\mu={\cal O}(1)$ it seems (although we have no
rigorous proof) that the magnetic instability is present independent
of the nature
of the gauge group (provided it has charged states in the
perturbative spectrum).

The behavior above should be compared to the field theory behavior
\be
E^2=p_3^2+M^2+eH(2n+1-gS)\label{class}
\ee
In (\ref{class}) we have an instability provided there is a particle
with $gS\geq 1$. Then the theory is unstable for
\be
|H|\geq {M^2\over |e|(gS-1)}
\ee
where $M$ is the mass of the particle (or the mass gap).
However there is no restauration of stability for large values of
$H$.
This happens in string theory due to the backreaction of gravity.
There is also another difference. In field theory, $H_{crit}\sim
\mu^2$
while in string theory $H_{crit}\sim \mu M_{str}$ where we denoted by
$\mu$ the mass gap in both cases and $M^2_{str}=1/\alpha'$.
We should also note that in a classically gapless theory like
unbroken Yang-Mills we obtain that the trivial ground state is
unstable even for infinitesimal
magnetic fields. This a tree level indication that this is not the
correct ground state of the theory, which of course can be verified
at one-loop where one learns that the coupling is strong in the IR
and the theory probably confines  and has a mass gap.

Let us now study the gravitational perturbation. Using (\ref{gener2})
the mass formula is (in analogy with (\ref{k11})
\be
M^2_{L}=-{1\over 2}+{\q^2\over 2}+{1\over
2}\sum_{i=1}^{3}\q_{i}^2+{(j+1/2)^2-(\q+I)^2\over
k+2}+E_{0}+\label{k22}
\ee
$$+{1+\sqrt{1+\rt^2}\over 2}\left[{(\q+I)\over \sqrt{k+2}}+{\rt\over
1+\sqrt{1+\rt^2}}{\bar I\over \sqrt{k}}\right]^2
$$
Introducing the $\sigma$-model variable
\be
\l=\sqrt{\rt +\sqrt{1+\rt^2}}\;\;\;,\;\;\;{1\over \l}=\sqrt{-\rt
+\sqrt{1+\rt^2}}
\ee
(\ref{k22}) becomes
\be
M^2_{L}=-{1\over 2}+{\q^2\over 2}+{1\over
2}\sum_{i=1}^{3}\q_{i}^2+{(j+1/2)^2-(\q+I)^2\over
k+2}+E_{0}+\label{k23}
\ee
$$+{1\over 4}\left[\left(\l+{1\over \l}\right){(\q+I)\over
\sqrt{k+2}}+\left(\l-{1\over \l}\right){\bar I\over
\sqrt{k}}\right]^2
$$
Only $|\q|=1$ and $j=|I|=0,1,\cdots,k/2$, can produce tachyonic
instabilities.
Due to the $\l\to 1/\l$ duality we will restrict ourselves to the
region $\l\leq 1$.

Thus, the condition for existence of tachyons is
\be
{1\over 4}\left[\left(\l+{1\over \l}\right){(\q+I)\over
\sqrt{k+2}}+\left(\l-{1\over \l}\right){\bar I\over
\sqrt{k}}\right]^2+
{\q^2-1\over 2}+{(|I|+1/2)^2-(\q+I)^2\over k+2}\leq 0\label{tach2}
\ee

Thus the state with quantum numbers $(I,\bar I)$  becomes tachyonic
when
\be
\l^2_{min}\leq \l^2 \leq \l^2_{max}
\ee
with
\be
\l^2_{max}={{\bar I^2\over k}-{I^2-1/2\over k+2}+\sqrt{{(I+3/4)\over
k+2}\left(
{\bar I^2\over k}-{(I+1/2)^2\over k+2}\right)}\over \left({I\over
\sqrt{k+2}}
+{\bar I\over \sqrt{k}}\right)^2}
\ee
\be
\l^2_{min}={{\bar I^2\over k}-{I^2-1/2\over k+2}-\sqrt{{(I+3/4)\over
k+2}\left(
{\bar I^2\over k}-{(I+1/2)^2\over k+2}\right)}\over \left({I\over
\sqrt{k+2}}
+{\bar I\over \sqrt{k}}\right)^2}
\ee
For large $k$, $\l_{max}$ approaches one, however at the same time
the
instability region shrinks to zero so that in the limit
$\l=1,k=\infty$ flat space is stable.

\section{Trace Formulae for small Magnetic Fields}
\setcounter{equation}{0}

In this section, we will treat the magnetic fields $F_{i}$, as well
as
the curvature perturbation $\rt$ as small, and we will derive trace
formulae for
averages of polynomials in this parameters, for applications to the
evaluation
of loop corrections to the appropriate coupling constants.

We will need a single magnetic field $F_{i}$ for each simple or U(1)
factor of the gauge group and a different normalization than the one
used in (\ref{gener})
\be
F_{i}\to \sqrt{k_{i}(k+2)}F_{i}\;\;\;,\;\;\;\rt \to \sqrt{k(k+2)}\rt
\ee
Then  (\ref{gener}) becomes
\be
\delta L_{0}=\delta \bar L_{0}=(\q+I)(\rt \bar I+F_{i}\pq_{i})+
\ee
$$+{-1+\sqrt{1+(k+2)(k_{i}F_{i}^2+k\R^2)}\over 2}\left[
{(\Q+I)^2\over k+2}+{(F_{i}\pq_{i}+\R\bar I)^2\over
k_{i}F_{i}^2+k\R^2}
\right]
$$
The first term is the linearized perturbation while the second is the
backreaction necessary for conformal and modular invariance.
The unperturbed partition function can be written as
\be
Tr[\exp[-2\pi \rm{Im}\tau (L_{0}+\bar L_{0})
+2\pi i\rm{Re}\tau (L_{0}-\bar L_{0})]]
\ee

Expanding the perturbed partition function in a power series in
$F_i,\R$
\begin{equation}
Z(\mu,F,\R)=\sum_{n_i,m=0}^{\infty}\prod_{i=1}^{n}F_i^{n_i}
\R^{m}Z_{n_i,m}(\mu)
\end{equation}
we can extract the integrated correlators $Z_{n_i,m}=\langle
\prod_{i=1}^{n}F_i^{n_i} R^m\rangle$ ($n$ is the number of simple
components of the gauge group).
\bs
\label{formu}
\be
\langle F_{i}\rangle =-4\pi {\rm Im}\tau \langle (\Q+I)\rangle\langle
\pq_{i}\rangle
\ee
\be
\langle \R\rangle =-4\pi {\rm Im}\tau \langle (\Q+I)\rangle\langle
\bar I\rangle
\ee
\be
\langle F_{i}^2\rangle =8\pi^2{\rm Im}\tau^2\left[ \langle
(\Q+I)^2\rangle
-{(k+2)\over 8\pi\im}\right]\left[\langle (\pq_{i})^2\rangle-{k_{i}
\over 8\pi{\rm Im}\tau}\right]-{k_{i}(k+2)\over 8}
\label{F2}
\ee
\be
\langle \R^2\rangle =8\pi^2{\rm Im}\tau^2\left[ \langle
(\Q+I)^2\rangle
-{k+2\over 8\pi\im}\right]\left[\langle (\bar I)^2\rangle-{k
\over 8\pi{\rm Im}\tau}\right]-{k(k+2)\over 8}
\ee
\be
\langle \R F_{i}\rangle = 16\pi^2{\rm Im}\tau^2 \langle \bar
I\pq_{i}\rangle
\left[\langle (\Q+I)^2\rangle-{k+2\over 8\pi{\rm Im}\tau}\right]
\ee
\be
\langle F_{i}F_{j}\rangle = 16\pi^2{\rm Im}\tau^2 \langle
\pq_{i}\pq_{j}\rangle
\left[\langle (\Q+I)^2\rangle-{k+2\over 8\pi{\rm Im}\tau}\right]
\ee
\es
where we should always remember that $k+2=1/\mu^2$.
We should note here that for torroidal $U(1)$ gauge fields there is
another natural basis in which $\langle \bar J_{i}(1)\bar
J_{j}(0)\rangle =G_{ij}/2$
where $G_{ij}$ is the constant metric of the torus.
Then the trace formula becomes
\be
\langle F_{i}F_{j}\rangle =8\pi^2{\rm Im}\tau^2\left[ \langle
(\Q+I)^2\rangle
-{(k+2)\over 8\pi\im}\right]\left[\langle (\pq_{i})^2\rangle-{G_{ij}
\over 8\pi{\rm Im}\tau}\right]-{G_{ij}(k+2)\over 8}
\label{F10}
\ee

For Supersymmetric ground states we have simplifications
\be
\langle F_{i}^2\rangle_{SUSY}=8\pi^2{\rm
Im\tau}^2\langle\Q^2\rangle\left[
\langle (\pq_{i})^2\rangle-{k_{i}\over 8\pi{\rm Im}\tau}\right]
\ee
\be
\langle \R^2\rangle_{SUSY}=8\pi^2{\rm
Im\tau}^2\langle\Q^2\rangle\left[
\langle (\bar I)^2\rangle-{k\over 8\pi{\rm Im}\tau}\right]
\ee

Renormalizations of higher terms can be easily computed.
We give here the expression for an $F_{i}^4$ term,

$$
\langle F_{i}^4\rangle={(4\pi{\rm Im}\tau)^4\over 24}\langle\left[
(\Q+I)^4  \pq_{i}^4-{3\over 4\pi{\rm Im}\tau}(\Q+I)^2\pq_{i}^2
\left((k_{i}(\Q+I)^2+\right.\right.
$$
\be
+\left.(k+2)\pq_{i}^2\right)
+{3\over 4(4\pi{\rm
Im}\tau)^2}\left[k_{i}(\Q+I)^2+(k+2)\pq_{i}^2\right]^2-\label{F4}
\ee
$$
-\left.{3k_{i}(k+2)\over 2(4\pi{\rm
Im}\tau)^3}\left[[k_{i}(\Q+I)^2+(k+2)\pq_{i}^2\right]
\right]\rangle
$$

The charge $\Q$ in the above formulae acts on the helicity
$\vartheta$-function
$\vartheta\left[^{\alpha}_{\beta}\right](\tau,v)$ as
differentiation with respect to $v$ divided by $2\pi i$.
The charges $\pq_{i}$ act also as $v$ derivatives on the respective
characters of the current algebra.
$I,\bar I$ act on the level-$k$ $\vartheta$-function present in
$SO(3)_{k/2}$ partition function (due to the parafermionic
decomposition).

\section{One-loop Corrections to the Coupling Constants}
\setcounter{equation}{0}

We now focus on  the one-loop correction to the
gauge
couplings.
Bearing anomalous U(1)'s we can immediately see from (\ref{formu})
that
$\langle F_{i}\rangle =0$ and $\langle F_{i}F_{j}\rangle =0$ for
$i\not= j$.
The conventionally normalized one-loop correction is
\be
{16\pi^2\over g^2_i}|_{1-loop}=-{1\over (2\pi)^2}\int_{\cal
F}{d^2\t\over \im^2}\langle F_{i}^2\rangle
\label{121}
\ee

Putting everything together we obtain

$${16\pi^2\over g_{i}^2}|_{1-loop} =-{i\over \pi^2 V(\mu)}\int_{\cal
F}{d^2\t
\over \im |\eta|^4}
\sum_{a,b=0}^{1}\left[X'(\mu)\partial_{\t}\left({\th[^a_b]\over
\eta}\right)+{1\over 6\mu^2}
\dot X'(\mu){\th[^a_b]\over \eta}\right]\times
$$
\be
\times Tr^{I}_{a,b}\left[
\langle \pq_{i}^2\rangle-{k_{i}\over 8\pi{\rm Im}\tau}\right]-
-{k_{i}\over 64\pi^3\mu^2 V(\mu)}\int_{\cal F}{d^2\t\over \im^2}
X'(\mu)Z_{0}\label{35}
\ee
where dot stands for derivative with respect to $\t$ and
$Tr^{I}_{ab}$ stands for the trace in the $(^a_b)$ sector of the
internal CFT.
Eq. (\ref{35}) is valid also for non-supersymmetric ground states.

When we have $N\geq 1$ supersymmetry it simplifies to\footnote{This
formula appeared in \cite{trieste} in a slightly different notation.}
\be
{16\pi^2\over g_{i}^2}|^{SUSY}_{1-loop} =-{i\over \pi^2
V(\mu)}\int_{\cal F}{d^2\t
\over \im |\eta|^4}
\sum_{a,b=0}^{1}\left[X'(\mu){\partial_{\t}\th[^a_b]\over
\eta}\right]Tr^{I}_{a,b}\left[
\langle \pq_{i}^2\rangle-{k_{i}\over 8\pi{\rm Im}\tau}\right]
\label{511}
\ee
The general formula (\ref{35}) can be split in the following way
\be
{16\pi^2\over g_{i}^2}|_{1-loop}=I_{1}+I_{2}+I_{3}\label{40}
\ee

\be
I_{1}=-{ i\over \pi^2 V(\mu)}\int_{\cal F}{d^2\t\over
\im|\eta|^4}X'(\mu)\sum_{a,b=0}^{1}
 \partial_{\t}\left({\th[^a_b]\over
\eta}\right)Tr^{I}_{a,b}\left[\langle (\pq_{i}^2\rangle-{k_{i}\over
8\pi{\rm Im}\tau}\right]
\label{41}
\ee

\be
I_{2}=-{i\over 6\pi^2\mu^2V(\mu)}\int_{\cal F}{d^2\t\over
\im|\eta|^4}\dot
X'(\mu)\sum_{a,b}^{1}
{\th[^a_b]\over \eta}Tr^{I}_{a,b}\left[\langle
\pq_{i}^2\rangle-{k_{i}\over
8\pi{\rm Im}\tau}\right]
\label{42}
\ee

\be
I_{3}=-{k_{i}\over 64\pi^3\mu^2 V(\mu)}\int_{\cal F}{d^2\t\over
\im^2}
X'(\mu)Z_{0}\label{43}
\ee
All the integrands are separately modular invariant.
The universal term in $I_{1}$ is due to an axion tadpole. $I_{3}$ is
the contribution of a dilaton tadpole. $I_{2}$ are extra helicity
contributions
due to the curved background.
Moreover $I_{2},I_{3}$ have power IR divergences which reflect
quadratic divergences in the effective field theory.
$I_{2},I_{3}$ are zero for supersymmetric ground states due to the
vanishing of the sum of the helicity theta functions.

We will now analyse the contribution of the massless sector to the
one-loop corrections.
Since
\be
-{1\over i\pi }\partial_{\t}\left({\th[^a_b]\over \eta}\right)\to
(-1)^F\left({1\over 12}-\chi^2\right)
\label{444}
\ee
where $\chi$ is the helicity of a state,
we obtain
\be
I_{1}^{massless}=-{1\over \pi}Str\left[\pq_{i}^2\left({1\over
12}-\chi^2\right)\right]{\bf J}_{1}(\mu)
+{k_{i}\over 8\pi^2}Str\left[{1\over 12}-\chi^2\right]{\bf
J}_{2}(\mu)
\label{445}
\ee
\be
I_{2}^{massless}=-{1\over 12\pi^2\mu^2}Str[\pq^2_{i}]{\bf
J}_{2}(\mu)+{k_{i}\over 48\pi^3\mu^2}Str[{\bf 1}]{\bf J}_{3}(\mu)
\label{446}
\ee
\be
I_{3}^{massless}=-{k_{i}\over 64\pi^3\mu^2}Str[{\bf 1}]{\bf
J}_{3}(\mu)
\label{447}
\ee
Here
\be
{\bf J}_{n}\equiv {1\over V(\mu)}\int_{\cal F}{d^2\t\over
\im^n}X'(\mu)
\label{431}
\ee
which can be evaluated to be
\be
{\bf J}_{1}(\mu)=2\pi\log\mu^2 +2\pi(\log\pi+\gamma_{E}-3+{3\over
2}\log 3) +{\cal O}(e^{-{1\over \mu^2}})
\label{442}
\ee
\be
{\bf J}_{2}(\mu)=-{4\pi^2\over 3}(1+\mu^2)\;\;,\;\;
{\bf J}_{3}(\mu)=-\pi\log 3-{28\pi^3\over 15}\mu^4+{\cal
O}(e^{-{1\over \mu^2}})
\label{4431}
\ee
\def\m{\mu_{e}}

We would like now to describe the same calculation in the effective
field theory.

This calculation proceeds along the same lines as above taking into
account the following differences.

$\bullet$ Now the mass gap is $\m^2=1/k$ and $V(\m)=1/(8\pi\m^3)$.

$\bullet$ $\Gamma_{0}/V(\m)$ is given by the momentum mode part of
the stringy expression:
\be
{\Gamma_{0}\over V(\m)}=-4\m^3\partial_{\m^2}\sqrt{\im}\sum_{n\in Z}
e^{-\pi\im\m^2(2n+1)^2}
\label{450}
\ee

$\bullet$ There is an incomplete cancelation of the $1/8\pi\mu^2\im$
piece in
(\ref{F2}). What remains is $1/4\pi\im$.

$\bullet$ The integral over $\im$ is done from $0$ to $\infty$. We
will have to regulate the UV divergences coming from the region of
integration around $t=0$.
 We will use for simplicity the Schwinger
regularization which amounts to integrating the parameter t in the
interval $[1/\pi\Lambda^2,\infty]$.

Then,
\be
{16\pi^2\over g_{i}^2}|_{1-loop}^{EFT}=L_{1}+L_{2}+L_{3}
\label{452}
\ee
where
\be
L_{1}=-{1\over \pi}Str\left[\pq_{i}^2\left({1\over
12}-\chi^2\right)\right]{\bf K}_{1}(\m)
+{k_{i}\over 8\pi^2}Str\left[{1\over 12}-\chi^2\right]{\bf K}_{2}(\m)
\label{453}
\ee
\be
L_{2}=-{1\over 4\pi^2}\left(1+{1\over 3\m^2}\right)Str[\pq^2_{i}]{\bf
K}_{2}(\m)+{k_{i}\over 16\pi^3}\left({1\over 2}+{1\over
3\m^2}\right)Str[{\bf 1}]{\bf K}_{3}(\m)
\label{455}
\ee
\be
L_{3}=-{k_{i}(1+2\m^2)\over 64\pi^3\m^2}Str[{\bf 1}]{\bf K}_{3}(\m)
\label{456}
\ee
and
\be
{\bf K}_{n}(\m)\equiv {1\over V(\m)}\int_{1\over
\pi\Lambda^2}^{\infty}{dt\over
t^n}\partial_{\m^2}\sqrt{t}\left[\th_{3}(it\m^2)-\th_{3}(4it\m^2)
\right]
\label{458}
\ee
The integrals can again be evaluated
\def\L{\Lambda}
\be
{\bf K}_{1}(\m,\L)=4\pi\log(\m/\L)+2\pi(\gamma_{E}-2)+
{\cal O}\left(e^{-\L^2/\m^2}\right)
\label{459}
\ee
and for $n>1$
\be
{\bf K}_{n}(\m,\L)={2\pi\L^{2n-2}\over
1-n}+8\pi^{2-n}(2n-3)(1-2^{2n-3})\Gamma(n-1)
\zeta(2n-2)\m^{2n-2}+
{\cal O}\left(e^{-\L^2/\m^2}\right)
\label{460}
\ee

In a similar fashion we can calculate the string one-loop correction
to the $R^2$ coupling with the result

\be
{1\over g_{R^2}^2}|_{1-loop} ={4\over \pi V(\mu)}\int_{\cal F}{d^2\t
\over \im|\eta|^4}
\sum_{a,b}^{1}\left[\partial_{\t}\left({\theta[^a_b]\over
\eta}\right)\left(\bar G_{2}-{1\over
6\mu^2}\partial_{\tb}\right)X'\right.
\label{48}\ee
$$\left.+{1\over 6\mu^2}{\theta[^a_b]\over \eta}\left(\bar
G_{2}-{1\over 6\mu^2}\partial_{\tb}\right)\partial_{\t}X'\right]+
{k(k+2)\over 16\pi V(\mu)}\int_{\cal F}{d^2\t\over \im^2}X'Z_{0}
$$
where
\be
\bar G_{2}\equiv \partial_{\tb}\log\bar\eta+{i\over 4\im}={1\over
2}\partial_{\tb}\log[\im\bar \eta^2]
\label{49}\ee

One-loop corrections to higher dimension operators can also be
computed.
We give here the result for $F_{\mu\nu}^4$ for $Z_{2}\times Z_{2}$
symmetric orbifold compactifications of the heterotic string. This
correction
gets contributions from all sectors including $N=4$ ones and it is
thus interesting for studying decompactification problems in string
theory.
The $N=4$ sector contribution to the $F_{\mu\nu}^4$  term for the
$E_{8}$ gauge group can be computed from (\ref{F4}) to be
\be
{1\over g^2_{F^4}}|^{E_{8}}_{1-loop}={1\over V(\mu)}\int_{\cal
F}{d^2\t\over \im^2}
X'(\mu)\prod_{i=1}^{3}\left[\im\Gamma_{2,2}(T_{i},U_{i})\right]
\sum_{a,b=0}^{1}{\bar\vartheta^8[^a_b]\over \bar \eta^{24}}\times
\label{56}
\ee
$$
\times
\sum_{\gamma,\delta=0}^{1}\bar\vartheta^7[^{\gamma}_{\delta}]\left(
{i\over \pi}\partial_{\tb}-{5\over 2\pi\im}\right)\left({i\over
\pi}\partial_{\tb}-{1\over
4\pi\im}\right)\bar\vartheta[^{\gamma}_{\delta}]
$$

\section{IR Flow Equations for Couplings}
\setcounter{equation}{0}

Once we have obtained the one-loop corrections to the coupling
constants, we can observe that they satisfy scaling type flows.
We will present here IR Flow Equations (IRFE) for differences of
gauge couplings.

The existence of IRFE is due to differential equations satisfied by
the
lattice sum of an arbitrary (d,d) lattice,
\be
Z_{d,d}={\rm Im}\tau^{d/2}\sum_{P_{L},P_{R}}e^{i\pi\tau
P^{2}_{L}/2-i\pi\bar\tau P_{R}^2/2}
\label{partition}
\ee
where
\be
P_{L,R}^2=\vec n G^{-1}\vec n+2\vec mBG^{-1}\vec n+\vec
m[G-BG^{-1}B]\vec m\pm 2\vec m \cdot \vec n
\label{momentum}
\ee
$\vec m,\vec n$ are integer d-dimensional vectors
and $G_{ij}$ ($ B_{ij}$) is a real symmetric (antisymmetric) matrix.
$Z_{d,d}$ is $O(d,d,Z)$ and modular invariant.
Moreover it satisfies the following second order differential
equation\footnote{The special case for $d=2$ of this equation was
noted and used in \cite{moduli,ant}.}:
\be
\left[ \left(G_{ij}{\partial\over \partial G_{ij}}+{1-d\over
2}\right)^2
+2G_{ik}
G_{jl}{\partial^2\over \partial B_{ij}\partial B_{kl}}
-{1\over 4}-4{\rm
Im}\tau^2{\partial^2\over \partial \tau\partial \bar
\tau}\right] Z_{d,d}=0\label{irfe}
\ee

The equation above involves also the modulus of the torus $\tau$.
Thus it can be used to convert the integrands for threshold
corrections to differences of coupling constants  into
total
derivatives on $\tau$-moduli space.
We will focus on gauge couplings of $Z_{2}\times Z_{2}$ orbifold
models.
To derive such an equation we start from the integral expressions of
such couplings (\ref{511})
to obtain
\be
\Delta_{AB}\equiv {16\pi^2\over g_{A}^{2}}-{16\pi^2\over g^{2}_{B}}=
-4\mu^3(b_{A}-b_{B})\int_{\cal F}{d^{2}\tau\over {\rm
Im}\tau^{2}}X'(\mu){\rm Im}\tau \Gamma_{2,2}(T,\bar T,U,\bar
U)\label{difere}
\ee
Eq. (\ref{difere}) does not apply to $U(1)$'s that can get enhanced
at special points of the moduli.
Using (\ref{irfe}) we obtain
\be
\left[\left(\mu{\partial \over \partial \mu}\right)^2-2\mu{\partial
\over \partial \mu}-16{\rm Im}T^2{\partial^2\over \partial T\partial
\bar T}\right]\Delta_{AB}=0\label{irfe2}
\ee
and we have also a similar one with  $T\to U$.
Note that for couplings that have a logarithmic behavior, the double
derivative of $\mu$ does not contribute.

We strongly believe that such equations also exist for single
coupling constants using appropriate differential equations for
$(d,d+n)$ lattices.

Notice that the IR scale $\mu$ plays the role of the RG scale
in the effective
field theory:
\be
{16\pi^2\over g_{A}^{2}(\mu)}={16\pi^2\over
g_{A}^2(M_{str})}+b_{A}\log
{M^2_{str}\over \mu^2}+ F_{A}(T_{i})+{\cal O}(\mu^2/M^2_{str})
\label{betaa}
\ee
where the moduli $T_{i}$ have been rescaled
by $M_{str}$ so they are dimensionless.
Second, the IRFE gives a scaling relation for the moduli dependent
corrections.
Such relations are very useful for determining the moduli dependence
of the threshold corrections.
We will illustrate below such a determination, applicable to the
$Z_{2}\times Z_{2}$ example described  above.

Using the expansion (\ref{betaa}) and applying the IRFE (\ref{irfe2})
we obtain
\be
{\rm Im}T^2{\partial^2\over \partial T\partial\bar
T}(F_{A}-F_{B})={1\over 4}(b_{A}-b_{B})\label{eq2}
\ee
and a similar one for $U$.
This non-homogeneous equation has been obtained in \cite{moduli,ant}.

Solving them we obtain
\be
F_{A}-F_{B}=(b_{B}-b_{A})\log[{\rm Im}T{\rm Im}U]+f(T,U)+g(T,\bar
U)+{\rm cc}
\ee
If at special points in moduli space, the extra massless states are
uncharged
with respect to the gauge groups appearing in (\ref{eq2}) then the
functions
$f$ and $g$ are non-singular inside moduli space.
In such a case duality invariance of the threshold corrections
implies
that
\be
F_{A}-F_{B}=(b_{B}-b_{A})\log[{\rm Im}T{\rm
Im}U|\eta(T)\eta(U)|^4]+{\rm constant}
\ee
This is the result obtained via direct calculation in \cite{moduli}.

It is thus obvious that the IRFE provides a powerful tool in
evaluating
general threshold corrections as manifestly duality invariant
functions of the moduli.

\section{Further Directions}
\setcounter{equation}{0}

Another set of important couplings that we have not explicitly
addressed in this paper are the Yukawa couplings.
Physical Yukawa couplings depend on the K\"ahler potential and the
superpotential.
The superpotential receives no perturbative contributions and thus
can be calculated at tree level.
The K\"ahler potential however does get renormalized so in order to
compute
the one-loop corrected Yukawa couplings we have to compute the
one-loop renormalization of the K\"ahler metric.
When the ground state has (spontaneously-broken) spacetime
supersymmetry the wavefunction renormalization of the scalars
$\phi_{i}$ are the same as those for their auxiliary fields $F_{i}$.
Thus we need to turn on non-trivial $F_{i}$, calculate their
effective action on the torus and pick the quadratic part
proportional to
$F_{i}\bar F_{\bar j}$.
This can be easily done using the techniques we developed in this
paper
since it turns out that the vertex operators \cite{atsen} for some
relevant $F$ fields are bilinears of left and right U(1) chiral
currents.

There are several other open problems that need to be addressed in
this
context.

The structure of higher loop corrections should be investigated.
A priori there is a potential problem, due to the dilaton, at higher
loops.
One would expect that since there is a region of spacetime where the
string coupling become arbitrarily strong, higher order computations
would be problematic. We think that this is not a problem in our
models, because
in Liouville models with N=4 superconformal symmetry (which is the
case we consider) there should be no divergence due to the dilaton at
higher loops.
However, this point need further study.
One should eventually analyze the validity of non-renormalization
theorems at higher loops \cite{ant} since they are of prime
importance for phenomenology.

The consequences of string threshold corrections for low energy
physics
should be studied in order to be able to make quantitative
predictions.

Finally, with respect to magnetic and gravitational instabilities,
more study is needed in order to draw model-independent conclusions
on the perturbative stability of string ground states. This is
important since it provides the only perturbative way to investigate
stability in a first quantized formulation.

\vskip 1cm

\centerline{\bf Acknowledgements}

We would like to thank the organizers of the Strings 95 conference
for giving us the opportunity to present our results.
One of us (C.K.) was  supported in part by EEC contracts
SC1$^*$-0394C and SC1$^*$-CT92-0789.


\begin{thebibliography}{9}


\bibitem{cand}
P.~Candelas, G.~Horowitz, A.~Strominger and E.~Witten, \np {\bf B258}
(1985) 46.

\bibitem{orbifold} L.J. Dixon, J. Harvey, C. Vafa and E. Witten, \np
{\bf B261} (1985) 678;\\ {\bf B274} (1986) 285;\\
K.S.~Narain, M.H.~Sarmadi and C.~Vafa, \np {\bf 288} (1987) 551.

\bibitem{nar}
K.S.~Narain, \np {\bf B169} (1986) 41;\\
K.S.~Narain, M.H.~Sarmadi and E.~Witten, \np {\bf B279} (1987) 369;

\bibitem{llsmap} W. Lerche, D. L\"ust and A.N. Schellekens, \pl {\bf
B181} (1986) 71;\\ \np {\bf B287} (1987) 477.

\bibitem{abk4d} H. Kawai, D.C. Lewellen and S.H.-H. Tye, \np {\bf
B288} (1987) 1;\\
I. Antoniadis, C. Bachas and C. Kounnas, \np {\bf
B289} (1987) 87.

\bibitem{llsmap} W. Lerche, D. L\"ust and A.N. Schellekens, \pl {\bf
B181} (1986) 71; \np {\bf B287} (1987) 477.

\bibitem{gepner} D. Gepner, \pl {\bf B199} (1987) 370;
\np {\bf B296} (1988) 757.

\bibitem{effcl}
E.~Witten, \pl {\bf B155} (1985) 151;\\
S.~Ferrara, C.~Kounnas and M.~Porrati, \pl {\bf B181} (1986) 263;\\
S.~Ferrara, L.~Girardello, C.~Kounnas and M.~Porrati, \pl {\bf B193}
(1987) 368;\\
I.~Antoniadis, J.~Ellis, E.~Floratos, D.V.~Nanopoulos and T.~Tomaras,
\pl {\bf B191} (1987) 96;\\
S.~Ferrara, L.~Girardello, C.~Kounnas and M.~Porrati, \pl {\bf B194}
(1987) 358;\\
M.~Cvetic, J.~Louis and B.~Ovrut, \pl {\bf B206} (1988) 227;\\
M.~Cvetic, J.~Molera and B.~Ovrut, \pr {\bf D40} (1989) 1140;\\
L.~Dixon, V.~Kaplunovsky and J.~Louis, \np {\bf B329} (1990) 27;\\
M. Cvetic, B. Ovrut and W. Sabra, hep-th/9502144.

\bibitem{moduli}
V.S.~Kaplunovsky, \np {\bf B307} (1988) 145;\\
L.J.~Dixon, V.S.~Kaplunovsky and J.~Louis, \np {\bf B355} (1991)
649.

\bibitem{dfkz} J.-P.~Derendinger, S.~Ferrara, C.~Kounnas and
F.~Zwirner, \np {\bf
B372} (1992) 145 and \pl {\bf B271} (1991) 307;\\
G.~Lopez Cardoso and B.A.~Ovrut, \np {\bf B369} (1992)351;\\
S. Ferrara, C. Kounnas, D. L\"ust and F. Zwirner, \np {\bf B365}
(1991) 431.


\bibitem{ant} I. Antoniadis, K. Narain and T. Taylor, \pl {\bf B276}
(1991) 37;\\
I. Antoniadis, E. Gava and K. Narain \pl {\bf B283} (1992) 209;
\np {\bf B383} (1992) 93;\\
I. Antoniadis, E. Gava, K.S. Narain and T. Taylor, \np {\bf B407}
(1993) 706 (hep-th/9212045);
ibid. {\bf B413}  (1994) 162 (hep-th/9307158); ibid. Nucl. Phys. {\bf
B432} (1994) 187 (hep-th/9405024).

\bibitem{n4kounnas} C. Kounnas, \pl {\bf B321} (1994) 26;
Proceedings of the  International ``Lepton-Photon
Symposium and Europhysics Conference on High Energy
Physics", Geneva, 1991, Vol. 1, pp. 302-306;
Proceedings of the International Workshop on ``String Theory, Quantum
Gravity and Unification of Fundamental Interactions", Rome, 21-26
September 1992.

\bibitem{ki} E. Kiritsis, in the Proceeding of the International
Europhysics Conf. on HEP, Marseille, 1993, Eds. J. Carr and M.
Perrotet ( hep-th/9309064).

\bibitem{worm} I. Antoniadis, S. Ferrara and C. Kounnas \np {\bf
B421}
(1994) 343 (hep-th/9402073).

\bibitem{kktopol} E. Kiritsis and C. Kounnas, \pl {\bf B331} (1994)
51 (hep-th/9404092).

\bibitem{n2} P. Mayr and S. Stieberger, Nucl. Phys. {\bf B407} (1993)
725 (hep-th/9303017), Nucl. Phys. {\bf B412} (1994) 502
(hep-th/9304055), hep-th/9504129;\\
P. Mayr, H. Nilles and S. Stieberger, Phys. Lett. {\bf B317} (1993)
53 (hep-th/9307171);\\
G. Lopes Cardoso, D. L\"ust and B. Ovrut, Nucl. Phys. {\bf B436}
(1995) 65 (hep-th/9410056);\\
V. Kaplunovsky and J. Louis, Nucl. Phys. {\bf B422} (1994) 57
(hep-th/9402005), hep-th/9502077;\\
G. Lopes Cardoso, D. L\"ust and T. Mohaupt, hep-th/9412071;\\
B. de Wit, V. Kaplunovsky, J. Louis and D. L\"ust, hep-th/9504006;\\
I. Antoniadis, S. Ferrara, E. Gava, K. Narain and T. Taylor,
hep-th/9504034;\\
K. Dienes and A. Faraggi, hep-th/9505018, hep-th/9505046.

\bibitem{min} J. Minahan, Nucl. Phys. {\bf B298} (1988) 36.

\bibitem{aben} I. Antoniadis, C. Bachas, J. Ellis and  D. Nanopoulos,
Nucl. Phys. {\bf B328} (1989) 117.

\bibitem{magn} E. Kiritsis and C. Kounnas, CERN-TH/95-171.


\bibitem{trieste} E. Kiritsis and C. Kounnas, Nucl. Phys. {\bf B41}
[Proc. Sup.] (1995) 331, (hep-th/9410212);\\
Nucl. Phys. {\bf B442} (1995) 472,  (hep-th/9501020).

\bibitem{open} A. Abouelsaood, C. Callan, C. Nappi and S. Yost, Nucl.
Phys. {\bf B280} (1987) 599;\\
C. Bachas and M. Porrati, Phys. Lett. {\bf B296} (1992) 77
(hep-th/9209032);\\
K. Behrndt, Nucl. Phys. {\bf B414} (1994) 114 (hep-th/9304096);\\
S. Ferrara and M. Porrati, Mod. Phys. Lett. {\bf A8} (1993) 2497
(hep-th/9306048).

\bibitem{bk} I. Antoniadis C. Bachas and A. Sagnotti, Phys. Lett.
{\bf B235} (1990) 255;\\
C. Bachas and E. Kiritsis, Phys. Lett. {\bf B325} (1994)
103 (hep-th/9311185).

\bibitem{rt} J. Russo and A. Tseytlin, hep-th/9411099,
hep-th/9502038, hep-th/9506071;\\
A. Tseytlin, Phys. Lett. {\bf B346} (1995) 55 (hep-th/9411198).

\bibitem{bachas} C. Bachas, hep-th/9503030.

\bibitem{SS} S. Ferrara, C. Kounnas and M. Porrati, Nucl. Phys. {\bf
B304} (1988) 500; Phys. Lett. {\bf B197} (1987) 135; Phys. Lett. {\bf
B206} (1988) 25;\\
C. Kounnas and M. Porrati, Nucl. Phys. {\bf B310} (1988) 355.

\bibitem{atsen} J. Attick, L. Dixon and A. Sen, Nucl. Phys. {\bf
B292} (1987) 109.

\end{thebibliography}
\end{document}